\documentstyle[prl,aps]{revtex}
\draft

\newcommand{\ba}{\begin{eqnarray}}
\newcommand{\ea}{\end{eqnarray}}
\input{psfig}

\begin{document}

\title{Band structure from random interactions}

\author{R. Bijker$^1$ and A. Frank$^{1,2}$} 

\address{$^1$Instituto de Ciencias Nucleares, 
Universidad Nacional Aut\'onoma de M\'exico, 
Apartado Postal 70-543, 04510 M\'exico, D.F., M\'exico \\ 
$^2$Centro de Ciencias F{\'{\i}}sicas, 
Universidad Nacional Aut\'onoma de M\'exico, 
Apartado Postal 139-B, Cuernavaca, Morelos, M\'exico}

\date{November 10, 1999}

\maketitle

\begin{abstract}
The anharmonic vibrator and rotor regions in nuclei are investigated 
in the framework of the interacting boson model using an ensemble of 
random one- and two-body interactions. We find a predominance of 
$L^P=0^+$ ground states, as well as strong evidence for the occurrence 
of both vibrational and rotational band structures. This remarkable 
result suggests that such band structures represent a far more 
general (robust) property of the collective model space than is 
generally thought. 
\end{abstract} 


A recent analysis of experimental energy systematics of medium and 
heavy even-even nuclei suggests a tripartite classification of 
nuclear structure into seniority, anharmonic vibrator and rotor 
regions \cite{Casten,Zamfir}. Plots of the excitation energies of the 
yrast states with $L^P=4^+$ against $L^P=2^+$ show a characteristic 
slope for each region: 1.00, 2.00 and 3.33, respectively. 
In each of these three regimes, the energy systematics is extremely 
robust. Moreover, the transitions 
between different regions occur very rapidly, typically with the 
addition or removal of only one or two pairs of nucleons. 
The transition between the seniority region (either semimagic or 
nearly semimagic nuclei) and the anharmonic vibrator regime (either 
vibrational or $\gamma$ soft nuclei) was addressed in a simple 
schematic shell model calculation and attributed to the proton-neutron 
interaction \cite{BFP}. The empirical characteristics of the 
collective regime which consists of the anharmonic vibrator and 
the rotor regions, as well as the transition between them,  
have been studied \cite{ZC,Jolos} in the framework of the 
interacting boson model (IBM) \cite{IBM}. An analysis of phase 
transitions in the IBM \cite{GK,DSI} has shown that the collective 
region is characterized by two basic phases (spherical and deformed) 
with a sharp transition region, rather than a gradual softening  
which is traditionally associated with the onset of deformation 
in nuclei \cite{IZC,Dimitri}.

In a separate development, the characteristics of low-energy 
spectra of many-body even-even nuclear systems have been studied 
recently in the context of the nuclear shell model with random 
two-body interactions \cite{JBD,Johnson}. Despite the random nature 
of the interactions, the low-lying spectra still show surprisingly 
regular features, such as a predominance of $L^P=0^+$ ground states 
separated by an energy gap from the excited states, and the 
evidence of phonon vibrations. The occurrence of these pairing 
effects cannot be explained by the time-reversal symmetry of the 
random interactions \cite{random}. A subsequent analysis of the pair 
transfer amplitudes has shown that pairing is a robust feature 
of the general two-body nature of shell model interactions and the 
structure of the model space \cite{JBDT}. On the other hand, no 
evidence was found for rotational band structures. 

The existence of robust features in the low-lying spectra 
of medium and heavy even-even nuclei \cite{Casten,Zamfir} suggests 
that there exists an underlying simplicity of low-energy nuclear 
structure never before appreciated. In order to address this point 
we carry out 
a study of the systematics of collective levels in the framework 
of the IBM with random interactions. In an analysis of energies and 
quadrupole transitions we show that despite the random nature (both 
in size and sign) of the interaction terms, regular features 
characteristic of the anharmonic vibrator and rotor regions emerge. 
Our results imply that these features are, to a certain extent, 
independent of the specific character of the interaction, and 
probably arise from the two-body nature of the Hamiltonian 
and the structure of the collective model space. 

In the IBM, collective nuclei are described as a system of $N$ 
interacting monopole and quadrupole bosons. We consider the most 
general one- and two-body IBM Hamiltonian $H = H_1 + H_2$. The one- 
and two-body matrix elements are chosen independently using a 
Gaussian distribution of random numbers with zero mean and variances 
\ba
\left< H_{1,\alpha \alpha'}^2 \right> &=& 
v^2 \, (1+\delta_{\alpha \alpha'}) ~, 
\nonumber\\
\left< H_{2,\beta \beta'}^2 \right> &=& \frac{1}{(N-1)^2} \, 
v^2 \, (1+\delta_{\beta \beta'}) ~. 
\label{ensemble}
\ea
Since the matrix elements of $H_1$ and $H_2$ are proportional 
to $N$ and $N(N-1)$, respectively, 
we have introduced a relative scaling between the one- 
and two-body interaction terms of $1/(N-1)$. The coefficient 
$v^2$ is independent of the angular momentum and represents an 
overall energy scale. The ensemble defined by Eq.~(\ref{ensemble}) 
is similar, but not identical, to the Two-Body Random Ensemble 
of \cite{French}. In all calculations we take $N=16$ bosons and 
1000 runs. For each set of randomly generated one- and two-body 
matrix elements we calculate the entire energy spectrum and 
the $B(E2)$ values between the yrast states. 

\begin{figure}
\centerline{\hbox{
\psfig{figure=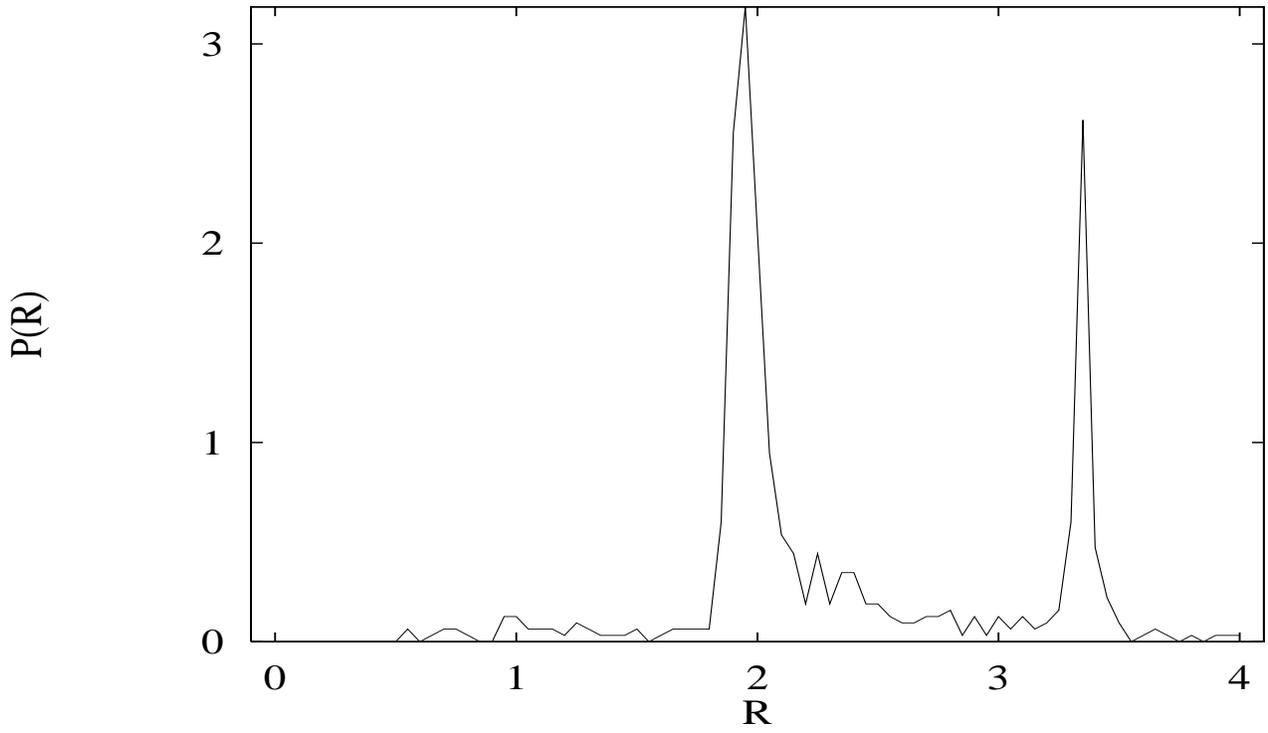,height=0.55\textwidth,width=1.0\textwidth} }}
\vspace{15pt}
\caption[]{Probability distribution $P(R)$ of the energy ratio 
$R=[E(4^+)-E(0^+)]/[E(2^+)-E(0^+)]$ with $\int P(R) dR = 1$ 
in the IBM with random one- and two-body interactions.}
\label{ratio}
\end{figure}

Just as in the case of the nuclear shell model \cite{JBD}, we 
find a predominance (63.4 $\%$) of $L^P=0^+$ ground states; 
in 13.8 $\%$ of the cases the ground state has $L^P=2^+$, 
and in 16.7$\%$ it has the maximum value of the angular momentum 
$L^P=32^+$. For the cases with a $L^P=0^+$ ground state 
we have calculated the probability distribution of the 
energy ratio $R=[E(4^+)-E(0^+)]/[E(2^+)-E(0^+)]$.  
Fig.~\ref{ratio} shows a remarkable result: the probability 
distribution $P(R)$ has two very pronounced peaks, one at 
$R \sim 1.95$ and a narrower one at $R \sim 3.35$. These values 
correspond almost exactly to the harmonic vibrator and rotor 
values (see the results for the $U(5)$ and $SU(3)$ limits 
in Table~\ref{BE2}). No such peak is observed for the $\gamma$ 
unstable or deformed oscillator case ($SO(6)$ limit). 

\begin{figure}
\centerline{\hbox{
\psfig{figure=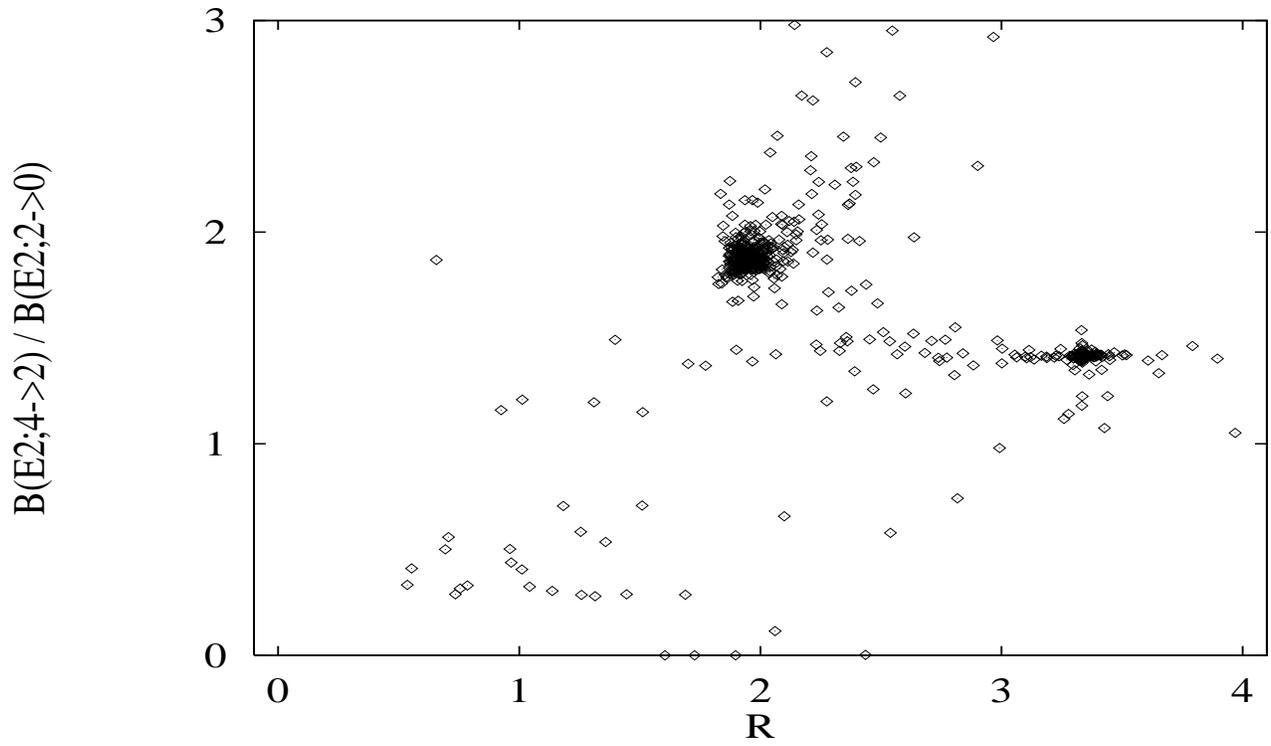,height=0.55\textwidth,width=1.0\textwidth} }}
\vspace{15pt}
\caption[]{Correlation between ratios of $B(E2)$ values and energies 
in the IBM with random one- and two-body interactions.}
\label{corr}
\end{figure}

Energies by themselves are not sufficient to decide whether 
or not there exists band structure. Levels belonging to a 
collective band are connected by strong electromagnetic 
transitions. In Fig.~\ref{corr} we show a correlation plot between 
the ratio of $B(E2)$ values for the $4^+ \rightarrow 2^+$ and 
$2^+ \rightarrow 0^+$ transitions and the energy ratio R. For  
the $B(E2)$ values we use the quadrupole operator
\ba
\hat Q_{\mu}(\chi) &=& ( s^{\dagger} \tilde{d} 
+ d^{\dagger} s)^{(2)}_{\mu} 
+ \chi \, (d^{\dagger} \tilde{d})^{(2)}_{\mu} ~, 
\label{qop}
\ea
with $\chi=-\sqrt{7}/2$. 
For completeness, in Table~\ref{BE2} we show the results for the 
three symmetry limits of the IBM \cite{IBM}. In the large $N$ limit, 
the ratio of $B(E2)$ values is 2 for the harmonic oscillator 
($U(5)$ limit) and 10/7 both for the deformed oscillator ($SO(6)$ 
limit) and the rotor ($SU(3)$ limit). There is a strong correlation 
between the first peak in the energy ratio and the vibrator value 
for the ratio of $B(E2)$ values (the concentration of points in this 
region corresponds to about 50 $\%$ of all cases), as well as for 
the second peak and the rotor value (about 25 $\%$ of all cases).  
For the region $2.3 \stackrel{<}{\sim} R \stackrel{<}{\sim} 3.0$ 
one can see a concentration of points around the value 1.4, 
which reflects the transition between the deformed oscillator and 
the rotor limits (see Table~\ref{BE2}). Calculations for 
different values of the number of bosons $N$ show the same results. 

\begin{figure}
\centerline{\hbox{
\psfig{figure=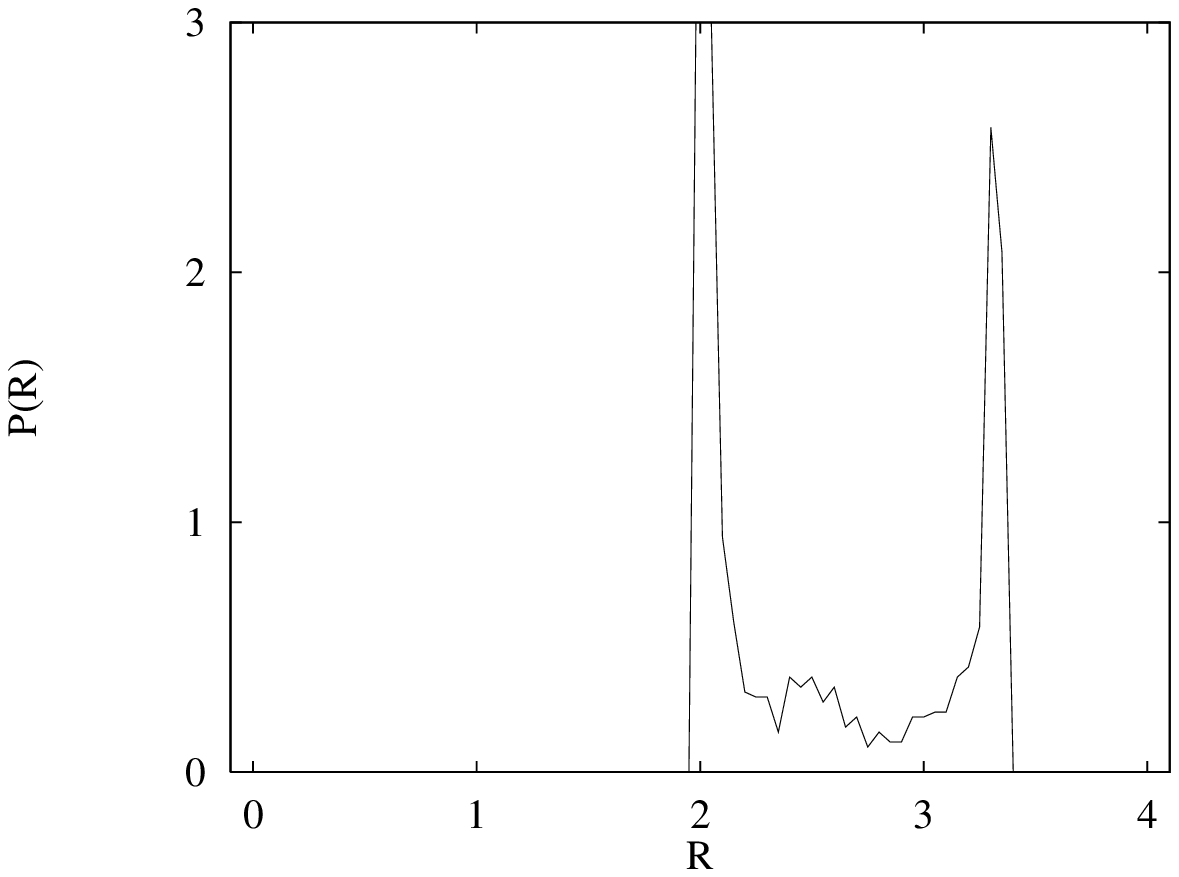,height=0.55\textwidth,width=1.0\textwidth} }}
\vspace{15pt}
\caption[]{As Fig.~\protect\ref{ratio}, but 
in the consistent Q-formulation of the IBM.}
\label{cqf1}
\end{figure}

\begin{figure}
\centerline{\hbox{
\psfig{figure=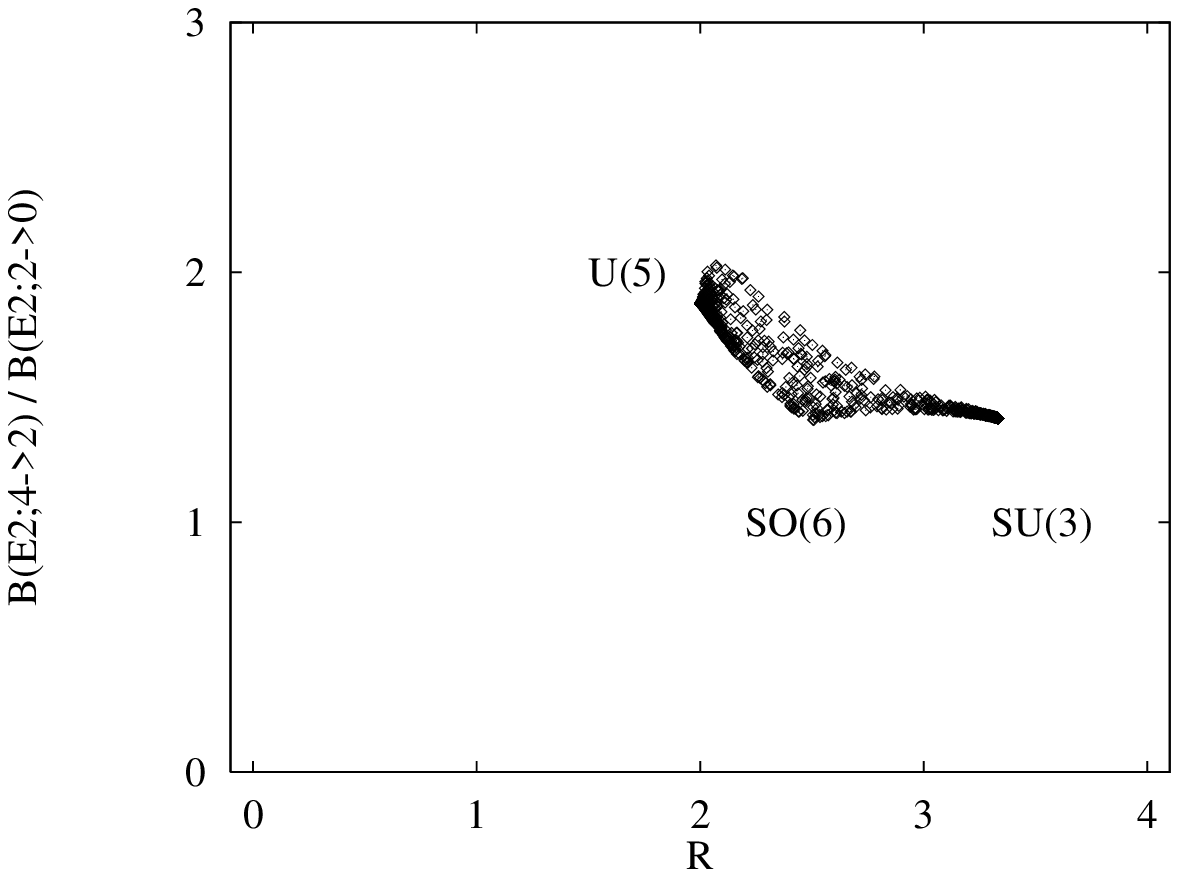,height=0.55\textwidth,width=1.0\textwidth} }}
\vspace{15pt}
\caption[]{As Fig.~\protect\ref{corr}, but 
in the consistent Q-formulation of the IBM.}
\label{cqf2}
\end{figure}

Despite the randomness of the interactions these results constitute 
strong evidence for the occurrence of both vibrational and rotational 
band structures. We have repeated the calculations for different 
values of the number of bosons $N$ and find the same results. 
Since the results presented in Figs.~\ref{ratio} and \ref{corr} 
were obtained with random interactions, with no restriction 
on the sign or size of the one- and two-body matrix elements,  
it is of interest to compare them with a calculation in which the 
parameters are restricted to the `physically' allowed region. 
To this end we consider the consistent Q-formulation \cite{CQF} 
which uses the same form for the quadrupole operator, Eq.~(\ref{qop}), 
{\it i.e.} with the same value of $\chi$, for the $E2$ operator and 
the Hamiltonian 
\ba
H &=& \epsilon \, \hat n_d 
- \kappa \, \hat Q(\chi) \cdot \hat Q(\chi) ~. 
\label{hcqf}
\ea
The parameters $\epsilon$ and $\kappa$ are restricted to be positive, 
whereas $\chi$ can be either positive or negative 
$-\sqrt{7}/2 \leq \chi \leq \sqrt{7}/2$. The properties of the 
Hamiltonian of Eq.~(\ref{hcqf}) can be investigated by taking 
the scaled parameters $\eta=\epsilon/[\epsilon+4\kappa(N-1)]$ and 
$\bar{\chi}=2\chi/\sqrt{7}$ randomly on the intervals 
$0 \leq \eta \leq 1$ and $-1 \leq \bar{\chi} \leq 1$ (these 
coefficients have been used as control parameters in a study of 
phase transitions in the IBM \cite{DSI,IZC}).  
In Figs.~\ref{cqf1} and \ref{cqf2} we show the corresponding 
probability distribution and correlation plot 
for the consistent Q-formulation of the IBM 
with realistic interactions. Although in this case the points 
are concentrated in a smaller region of the plot than before, 
the results show the same qualititative behavior as for the IBM 
with random one- and two-body interactions. In Fig.~\ref{cqf2} 
we have identified each of the dynamical symmetries 
of the IBM (and the transitions between them). 
There is a large overlap between the regions with the highest 
concentration of points in Figs.~\ref{corr} and \ref{cqf2}. 

In conclusion, we have studied the IBM using random ensembles 
of one- and two-body Hamiltonians. It was found that despite the 
randomness of the interactions the ground state has $L^P=0^+$ 
in 63.4 $\%$ of the cases. For this subset, the analysis of both 
energies and quadrupole transitions shows strong evidence for the 
occurrence of both vibrational and rotational band structure. These 
features arise from a much wider class of Hamiltonians than are 
generally considered to be `realistic'. This suggests that these band 
structures arise, at least in part, as a consequence of the 
one- and two-body nature of the interactions and the structure 
of the collective model space, and hence represent a far more 
general and robust property of collective Hamiltonians than is commonly 
thought. This is in qualitative agreement with the empirical 
observations of robust features in the low-lying spectra of 
medium and heavy even-even nuclei \cite{Casten,Zamfir}. 

A similar situation has been observed in the context of the nuclear 
shell model with respect to the pairing properties \cite{JBD,JBDT} 
which were formerly exclusively attributed to the particular form of the 
nucleon-nucleon force. On the other hand, the random IBM Hamiltonians 
studied in this Letter display not only vibrational-like phonon 
collectivity but, in contrast to the results in \cite{JBD,JBDT}, 
also imply the emergence of rotational bands. The IBM is based on 
the assumption that low-lying collective excitations in nuclei 
can be described as a system of interacting monopole and quadrupole 
bosons, which in turn are associated with generalized pairs of 
like-nucleons with angular momentum $L=0$ and $L=2$.  
It would be very interesting to establish whether rotational 
features can also arise from ensembles of random interactions in 
the nuclear shell model, if appropriate (minimal) restrictions 
are imposed on the parameter space. 

It is a pleasure to thank Stuart Pittel and Rick Casten for 
interesting discussions. This work was supported in part by 
DGAPA-UNAM under project IN101997.

\begin{table}
\caption[]{Energies of $B(E2)$ values in the dynamical symmetry 
limits of the IBM \protect\cite{IBM}. In the $U(5)$ and $SO(6)$ 
limits we show the result for the leading order contribution to the 
rotational spectra.} 
\label{BE2}
\vspace{15pt}
\begin{tabular}{ccc}
& \\
& $\frac{E(4^+)-E(0^+)}{E(2^+)-E(0^+)}$  
& $\frac{B(E2;4^+ \rightarrow 2^+)}{B(E2;2^+ \rightarrow 0^+)}$ \\
& \\
\hline
& \\
$U(5)$  & $2$ & $\frac{2(N-1)}{N}$ \\
$SO(6)$ & $\frac{5}{2}$ 
& $\frac{10(N-1)(N+5)}{7N(N+4)}$ \\
$SU(3)$ & $\frac{10}{3}$ 
& $\frac{10(N-1)(2N+5)}{7N(2N+3)}$ \\
& \\
\end{tabular}
\end{table}

\end{document}